# A Fast Quantum-safe Asymmetric Cryptosystem Using Extra Superincreasing Sequences

Shenghui Su [1, 4], Jianhua Zheng [2], and Shuwang Lü [3, 5]

[1] College of Computers, Nanjing Univ. of Aeronautics & Astronautics, Nanjing 211106, PRC
[2] School of Network Securities, Information Engineering University, Zhengzhou 450001, PRC
[3] School of Computers, Univ. of Chinese Academy of Sciences, Beijing 100039, PRC
[4] Public Security Innovation Center, Nanjing Univ. of Science and Technology, Nanjing 210094, PRC
[5] Laboratory of Computational Complexity, BFID Corporation, Beijing 100098, PRC
{Corresponding email: reesse@126.com / Received: Sep 05, 2017, Revised: Oct 01, 2017}

**Abstract**: This paper gives the definitions of an extra superincreasing sequence and an anomalous subset sum, and proposes a fast quantum-safe asymmetric cryptosystem called JUOAN2. The new cryptosystem is based on an additive multivariate permutation problem (AMPP) and an anomalous subset sum problem (ASSP) which parallel a multivariate polynomial problem and a shortest vector problem respectively, and composed of a key generator, an encryption algorithm, and a decryption algorithm. The authors analyze the security of the new cryptosystem against the Shamir minima accumulation point attack and the LLL lattice basis reduction attack, and prove it to be semantically secure (namely IND-CPA) on the assumption that AMPP and ASSP have no subexponential time solutions. Particularly, the analysis shows that the new cryptosystem has the potential to be resistant to quantum computing attack, and is especially suitable to the secret communication between two mobile terminals in maneuvering field operations under any weather. At last, an example explaining the correctness of the new cryptosystem is given.

## 1  Introduction

The MH knapsack asymmetric cryptosystem [1] is frangible individually to the Shamir minima accumulation point finding which attacks a public key [2] and to the LLL lattice basis reduction which attacks a ciphertext [3][4][5], but it delivers crucial enlightenment to succeeding designers of fast asymmetric cryptosystems. Besides, the REESSE1+ and JUOAN asymmetric cryptosystems deliver recent enlightenment to designers of quantum-resistant cryptosystems based on multivariate problems and lattice problems [6][7].

In this paper, we intend to design a fast quantum-safe asymmetric cryptosystem which will be based on a multivariate problem as well as a lattice problem, and will use an extra superincreasing sequence as well as an anomalous subset sum.

Throughout the paper, unless otherwise specified, the sign % means "modulo", $\lg x$ denotes the logarithm of $x$ to the base 2, $\neg b_i$ does the negative value of a bit $b_i$, and $\|x\|$ does the order of an element $x$ % $M$. Without ambiguity, "% $M$" is usually omitted in expressions.

## 2  Several Definitions and Related Properties

**Definition 1**: For $n$ positive integers $A_1, A_2, \ldots,$ and $A_n$, if $A_2 > A_1 + 1$ and every $A_i$ ($i > 2$) satisfies
$$A_i > \sum_{j=1}^{i-1} (i-j) A_j,$$
then this series of integers is called an extra superincreasing sequence, denoted by $\{A_1, \ldots, A_n\}$, and shortly $\{A_i\}$.

For example, {1, 3, 8, 21, 54, 139, 367, 960} is an extra superincreasing sequence.

**Property 1:** Assume that $\{A_1, \ldots, A_n\}$ is an extra superincreasing sequence. Then, for $i > 1$ and any positive integer $k$, there exists $(k + 1)A_i > \sum_{j=1}^{i-1}(k + i - j)A_j$.

***Definition 2:*** Assume that $b_1\ldots b_n \in \{0, 1\}^n$ is a plaintext block, $r_1\ldots r_n \in \{0, 1\}^n$ is a random noise, $M$ is an integral modulus, and $\{C_1, \ldots, C_n \mid C_i \in [1, M-1]\}$ is an integer sequence (namely an ordered set). Then

$$\dot{S} \equiv \sum_{i=1}^{n}(b_i \oplus r_i)L_i C_i \ (\% \ M)$$

is called an anomalous subset sum, where $L_i = \sum_{j=i}^{n} b_j$.

Obviously, when $r_1\ldots r_n = 0$, $\dot{S}$ is simplified to $\sum_{i=1}^{n} b_i L_i C_i \ (\% \ M)$, and when $r_1\ldots r_n = 0$ and $L_i$ does not exist there, $\dot{S}$ degenerates to a subset sum $\sum_{i=1}^{n} b_i C_i \ (\% \ M)$.

Notice that in Definition 2, $\{C_1, \ldots, C_n\}$ is not necessarily an extra superincreasing sequence or a superincreasing sequence.

***Property 2:*** For any positive integer $m \leq n$, if randomly select $m$ elements from an extra superincreasing sequence $\{A_1, \ldots, A_n\}$, and construct a subsequence (namely an ordered subset) $\{A_{x_1}, \ldots, A_{x_m}\}$, then an anomalous subset sum $S = mA_{x_1} + (m-1)A_{x_2} + \ldots + A_{x_m}$ is uniquely determined, that is, the mapping from $S$ to $\{A_{x_1}, \ldots, A_{x_m}\}$ is one-to-one.

***Definition 3:*** In an asymmetric cryptosystem, the parameter $\ell(i)$ in the key transform is called the lever function, if it has the following features [6][7]:

- $\ell(.)$ is an injection from integers to integers, its domain is $[1, n]$, and codomain $[1, M]$. Let $Ł_n$ represent the collection of all injections from the domain to the codomain, then $\ell(.) \in Ł_n$ and $|Ł_n| \geq A_n^n = n(n-1)\ldots 1$.
- The mapping between $i$ and $\ell(i)$ is established randomly without an analytical formula, so every time a public key is generated, the function $\ell(.)$ is distinct.
- There does not exist any dominant or special mapping from $\ell(.)$ to a public key.
- An attacker has to consider all the arrangements of the sequence $\{\ell(1), \ldots, \ell(n)\}$ when extracting a related private key from a public key. Thus, if $n$ is large enough, it is infeasible for the attacker to search the arrangements exhaustively.
- A receiver owning a private key only needs to consider the accumulative sum of $n$ elements in $\{\ell(1), \ldots, \ell(n)\}$ when recovering a related plaintext from a ciphertext. Thus the time complexity of decryption is polynomial in $n$, and the decryption is feasible.

Obviously, there is the large amount of calculation on $\ell(.)$ at "a public terminal", and the small amount of calculation on $\ell(.)$ at "a private terminal".

***Property 3 (Indeterminacy of $\ell(.)$):*** Let $C_i \equiv (A_i + W\ell(i))\delta \ (\% \ M)$ with $\delta = 1$, $M$ an integral modulus, and $\ell(i) \in \Omega = \{e_1, \ldots, e_n \mid e_i \leq 2n\}$ for $i = 1, \ldots, n$. Then $\forall \ W \ (\|W\| \neq M) \in (1, M)$ and $\forall \ x, y, z \ (x \neq y \neq z) \in [1, n]$,

① when $\ell(x) + \ell(y) = \ell(z)$, there is $\ell(x) + \|W\| + \ell(y) + \|W\| \neq \ell(z) + \|W\| \ (\% \ M)$;

② when $\ell(x) + \ell(y) \neq \ell(z)$, there always exist

$$C_x \equiv A'_x + W'\ell'(x) \ (\% \ M), \ C_y \equiv A'_y + W'\ell'(y) \ (\% \ M), \ \text{and} \ C_z \equiv A'_z + W'\ell'(z) \ (\% \ M)$$

such that $\ell'(x) + \ell'(y) \equiv \ell'(z) \ (\% \ M)$ with the constraint $A'_z <$ or $\approx M / 3$.

## 3 New Asymmetric Cryptosystem

The new cryptographic scheme includes three algorithms separately for key pair, encryption, and decryption.

### 3.1 Key Generation Algorithm

Suppose that $n$ is the bit-length of a plaintext block or a symmetric key, and let $\tilde{n} = 3n / 2$.

This algorithm is employed by a third-party authority or a user who needs to hold a private key exclusively.

INPUT: an integer $\tilde{n}$.
S1: Randomly generate an extra superincreasing sequence $\{A_1, \ldots, A_{\tilde{n}}\}$.
S2: Find an integer $M$ making $M > \sum_{i=1}^{\tilde{n}} (\tilde{n} + 1 - i) A_i$ and $1.585\tilde{n} \leq \lceil \lg M \rceil \leq 2\tilde{n}$.
S3: Pick arbitrary integers $W, \delta \in [1, M - 1]$ such that $\gcd(\delta, M) = 1$.
   Calculate $-W$ by $W + (-W) \equiv 0 \ (\% M)$ and $\delta^{-1}$ by $\delta \delta^{-1} \equiv 1 \ (\% M)$.
S4: Produce pairwise distinct $\ell(1), \ldots, \ell(\tilde{n}) \in \Omega = \{e_1, \ldots, e_{\tilde{n}} \mid e_i \leq 2\tilde{n}\}$.
S5: Compute a sequence $\{C_1, \ldots, C_{\tilde{n}} \mid C_i \equiv (A_i + W\ell(i))\delta \ \% \ M\}$.
OUTPUT: a public key $(\{C_i\}, M)$; a private key $(\{A_i\}, -W, \delta^{-1}, M)$.

Notice that the lever values $\{\ell(1), \ldots, \ell(\tilde{n})\}$ is discarded, and a random permutation of the subscript integers $\{1, \ldots, \tilde{n}\}$ may be considered.

**Definition 4**: Given a public key $(\{C_1, \ldots, C_{\tilde{n}}\}, M)$, seeking a related private key $(\{A_1, \ldots, A_{\tilde{n}}\}, W, \delta)$ by $C_i \equiv (A_i + W\ell(i))\delta \ (\% \ M)$ with $\{A_i\}$ an extra superincreasing sequence and $\ell(i)$ from $\Omega = \{e_1, \ldots, e_{\tilde{n}} \mid e_i \leq 2\tilde{n}\}$ for $i = 1, \ldots, \tilde{n}$ is called an additive multivariate permutation problem (AMPP).

It is not difficult to see an additive multivariate permutation problem is a form of a multivariate polynomial problem. Notice that the parameter $\tilde{n}$ in Definition 4 may be substituted with $n$.

### 3.2 Encryption Algorithm

INPUT: an integer $\tilde{n}$;
   a public key $(\{C_1, \ldots, C_{\tilde{n}}\}, M)$;
   an $n$-bit plaintext block or symmetric key $b_1 \ldots b_n \neq 0$
S1: extend $b_1 \ldots b_n$ to $b_1 \ldots b_n b_{n+1} \ldots b_{\tilde{n}}$,
   where $b_{n+1} \ldots b_{\tilde{n}} \in \{0, 1\}^{n/2}$ is a random padding.
S2: Yield a random noise $r_1 \ldots r_{\tilde{n}} \in \{0, 1\}^{\tilde{n}}$.
S3: Set $\dot{S} \leftarrow 0, L \leftarrow 0, i \leftarrow \tilde{n}$.
S4: If $b_i = 1$ then do $L \leftarrow L + 1$ and $\dot{S} \leftarrow \dot{S} + L C_i \ \% \ M$;
   else if $r_i = 1$ then do $\dot{S} \leftarrow (\dot{S} + L C_i) \ \% \ M$.
S5: Let $i \leftarrow i - 1$.
S6: If $i \geq 1$ then goto S4; else prepare outputting.
OUTPUT: a ciphertext $\dot{S}$.

Apparently, $\dot{S}$ may be written as $\dot{S} \equiv \sum_{i=1}^{\tilde{n}} (b_i \oplus r_i) L_i C_i \ (\% \ M)$, where $L_i = \sum_{j=i}^{\tilde{n}} b_j$.

**Definition 5**: Given a public key $(\{C_1, \ldots, C_{\tilde{n}}\}, M)$ and a ciphertext $\dot{S}$, seeking a related plaintext $b_1 \ldots b_{\tilde{n}}$ by $\dot{S} \equiv \sum_{i=1}^{\tilde{n}} (b_i \oplus r_i) L_i C_i \ (\% \ M)$ with $L_i = \sum_{j=i}^{\tilde{n}} b_j$ and $b_i, r_i \in [0, 1]$ is called an anomalous subset sum problem (ASSP).

It should be noted that the parameter $\tilde{n}$ in Definition 5 may be substituted with $n$.

### 3.3 Decryption Algorithm

INPUT: an integer $\tilde{n}$;
   a private key $(\{A_1, \ldots, A_{\tilde{n}}\}, -W, \delta^{-1}, M)$;
   a ciphertext $\dot{S}$.
S1: Compute $\dot{S} \leftarrow \dot{S} \delta^{-1} \ \% \ M$.
S2: Compute $\dot{S} \leftarrow \dot{S} + (-W) \ \% \ M$.
S3: Set $b_1 \ldots b_{\tilde{n}} \leftarrow 0, \S \leftarrow \dot{S}, L \leftarrow 0, i \leftarrow \tilde{n}$.
S4: If $\S \geq (L + 1) A_i$ then do $b_i \leftarrow 1, L \leftarrow L + 1$ and $\S \leftarrow \S - L A_i$;
   else if $\S \geq L A_i$ then do $\S \leftarrow \S - L A_i$.
S5: Let $i \leftarrow i - 1$.

    If $i \geq 1$ and $\dot{S} \neq 0$ then goto S4.
S6: If $\dot{S} \neq 0$ then goto S2;
    else prepare outputting.
OUTPUT: a related plaintext $b_1…b_{\tilde{n}}$ containing the original plaintext $b_1…b_n$.
This algorithm can always terminate normally as long as $\dot{S}$ is a true ciphertext.

## 4 Preliminary Analysis of Security

The analysis of security of the new asymmetric cryptosystem should cover three aspects.

In this section, we replace the parameter $\tilde{n}$ with $n$ for writing convenience, which will not influence the correctness of analysis results.

### 4.1 Resistant to Shamir Minima Accumulation Point Attack

In the new asymmetric cryptosystem, the key transform is $C_i \equiv (A_i + W\ell(i))\delta \ (\% \ M)$ for $i = 1, …, n$ with $\{A_i\}$ an extra superincreasing sequence.

The above key transform is of a multivariate problem, and utterly different to the MH transform $c_i \equiv a_i W \ (\% \ M)$ for $i = 1, …, n$ with $\{a_i\}$ a superincreasing sequence. Therefore, the Shamir minima accumulation point attack [2] will be ineffective on public keys in the new cryptosystem.

### 4.2 Resistant to LLL Lattice Basis Reduction Attack

In the MH knapsack cryptosystem [1], a superincreasing sequence $\{a_1, …, a_n\}$ — $\{1, 2, …, 2^{n-1}\}$ for example, is a private key, a sequence $\{C_1, …, C_n \mid C_i \equiv a_i W \ (\% \ M)\}$ is a public key, a bit string $b_1…b_n \in \{0, 1\}^n$ is a plaintext block, and a subset sum $S = \sum_{i=1}^{n} b_i C_i \ \% \ M$ is a ciphertext.

Corresponding to the form of a subset sum, a related knapsack density is expressed as [3]
$$D = n / \lg \max_{1 \leq i \leq n} \{c_i\}$$
$$\approx n / \lg M.$$

When $D$ is less than 0.6463, and even 0.9408, a related plaintext $b_1…b_n$ can be recovered from a ciphertext $S$ through the LLL lattice basis reduction algorithm [3][4][5]. Therefore, the MH knapsack cryptosystem is insecure.

In the new asymmetric cryptosystem, an anomalous subset sum is
$$\dot{S} \equiv \sum_{i=1}^{n} (b_i \oplus r_i) L_i C_i \ (\% \ M) \text{ with } L_i = \sum_{j=i}^{n} b_j$$
which represents a ciphertext.

Corresponding to the form of an anomalous subset sum, a related knapsack density is
$$D = \sum_{i=1}^{n} \lg L_i / \lg M$$
$$= \lg (L_1 … L_n) / \lg M$$
$$= \lg n! / \lg M$$
$$\geq \lg n! / (2n).$$

It is should be noted that the maximal value of every $L_i$ on a bit string $b_1…b_n$ is $n - i + 1$, and thus the maximal bit-length of every $L_i$ is $\lg L_i = \lg(n - i + 1)$ which must be considered in the LLL lattice basis reduction algorithm.

When $n = 10$ at most, $D$ will be larger than 1. Hence, the LLL lattice basis reduction attack is ineffective on ciphertexts in the new cryptosystem.

### 4.3 Having Potential to Be Resistant to Quantum Computing Attack

It is well known that asymmetric cryptosystems based on integer factorization problems, discrete logarithm problems, or elliptic curve discrete logarithm problems — RSA, ElGamal, and ECC for example will be insecure on quantum computers [8][9].

***Definition 6***: Let $\{C_1, \ldots, C_n\}$ be a sequence of which every element is chosen uniformly within the domain $[1, M-1]$, and $b_1 \ldots b_n$ be a plaintext block which is chosen uniformly within the domain $\{0, 1\}^n$, and then let $S = \sum_{i=1}^{n} b_i C_i \% M$. Given $\{C_1, \ldots, C_n\}$ and $S$, finding a bit string $x_1 \ldots x_n \in \{0, 1\}^n$ such that $S = \sum_{i=1}^{n} x_i C_i \% M$ is called the average case subset sum problem [10][11][12].

O. Regev shows a corollary that if there exists an algorithm that solves a non-negligible part of instances with modulus $N$ of the average case subset sum problem (SSP), then there exists a quantum algorithm for the $\Theta(n^{2.5})$ unique shortest vector problem (SVP) [12], where $n$ is the dimension of a lattice. The corollary indicates the unique SVP will be secure on quantum computers as long as a polynomial time algorithm for a non-negligible part of instances of the average case SSP is not found. The unique SVP as a lattice problem has important cryptographic applications — a cryptosystem by Regev for example [13]. Consequently, lattice-based cryptography is on the list of quantum-resistant cryptographic techniques which is given by NIST [14]. In addition, multivariate polynomial cryptography is included by the list [14].

The new asymmetric cryptosystem is based on the hardnesses of a multivariate permutation problem and an anomalous subset sum problem which are separately equivalent to a multivariate polynomial problem and a shortest vector problem, and naturally, it has the potential to resist quantum computing attack.

## 5  Proof of Semantic Security

***Definition 7***: A cryptosystem is said to be semantically secure if an adversary who knows the encryption algorithm of the cryptosystem and is in possession of a ciphertext is unable to determine any information about the related plaintext [15].

It is subsequently demonstrated that semantic security is *equivalent* to another definition of security called ciphertext indistinguishability [16]. If a cryptosystem has the property of indistinguishability, then an adversary will be unable to distinguish a pair of ciphertexts based on the two plaintexts encrypted by a challenger.

A chosen plaintext attack (CPA) is an attack model for cryptanalysis which presumes that the attacker has the capability to choose arbitrary plaintexts to be encrypted and obtain the corresponding ciphertexts that are expected to decrease the security of a cryptosystem [17].

***Definition 8***: A cryptosystem is said to be IND-CPA (indistinguishable under chosen plaintext attack), namely semantically secure against chosen plaintext attack, if the adversary cannot determine which of the two plaintexts was chosen by a challenger, with probability significantly greater than 1/2, where 1/2 means the success rate of random guessing [17][18].

For a probabilistic asymmetric cryptosystem based on computational security, indistinguishability under chosen plaintext attack is illuminated by a game between an adversary and a challenger, where the adversary is regarded as a probabilistic polynomial time Turing machine, which means that it must complete the game and output a guess within a polynomial number of steps.

Notice that for the new asymmetric cryptosystem, an adversary may be also regarded as a probabilistic subexponential time Turing machine since no subexponential time solution to AMPP or ASSP is found so far.

***Theorem 1***: The new asymmetric cryptosystem is indistinguishable under chosen plaintext attack (IND-CPA) on the assumption that AMPP and ASSP cannot be solved in subexponential time.

*Proof*:

Let $E(k_p, \underline{m})$ represents the encryption of a message (plaintext) $\underline{m}$ under the public key $k_p$.

A game between an adversary and a challenger is given as follows.

① The challenger calls the key generation algorithm with the parameter $\tilde{n}$, obtains a key pair $(k_p, k_s)$, publishes $k_p = (\{C_1, \ldots, C_{\tilde{n}}\}, M)$ to the adversary, and retains $k_s$ for himself.

② The adversary may perform any number of encryptions or other compatible operations, which will give the adversary a weak advantage over tossing a coin.

③ Eventually, the adversary chooses any two distinct $n$-bit plaintexts ($\underline{m}_0$, $\underline{m}_1$), and submits them to the challenger.

④ The challenger selects a bit $x \in [0, 1]$ uniformly at random, and sends the challenge ciphertext
$$\dot{S} = E(k_p, \underline{m}_x) = \sum_{i=1}^{\tilde{n}} (b_i \oplus r_i) L_i C_i \% M$$
back to the adversary, where $\underline{m}_x = b_1 \ldots b_n$, and $b_1 \ldots b_{\tilde{n}} = b_1 \ldots b_n b_{n+1} \ldots b_{\tilde{n}}$ with $b_{n+1} \ldots b_{\tilde{n}}$ being a random padding.

⑤ The adversary is free to perform any number of additional computations or encryptions. Finally, it outputs a guess for the value of $x$. Therefore, it is needed to analyze the probability of hitting $x$.

Because the intricacies AMPP and ASSP have no subexponential time solutions, neither can the adversary acquire a private key from a public key $(\{C_1, \ldots, C_{\tilde{n}}\}, M)$, nor can directly solve the equation
$$\dot{S} \equiv \sum_{i=1}^{\tilde{n}} (b_i \oplus r_i) L_i C_i \ (\% \ M)$$
for $b_1 \ldots b_{\tilde{n}} = b_1 \ldots b_n b_{n+1} \ldots b_{\tilde{n}}$ which contains $\underline{m}_x$.

It is known from the encryption algorithm that one identical plaintext $b_1 \ldots b_n$ may be mapped to $2^{n/2} \underline{r}(L_1)$ different ciphertexts, where the function $\underline{r}(L_1) > 1$ is brought forth by a random bit noise $r_1 \ldots r_{\tilde{n}}$. It will take the running time of $O(2^{n/2} \underline{r}(L_1))$ encryption operations to verify all the possible ciphertexts of a plaintext.

Thus the probability that the adversary hits $x$ by guessing is only
$$\beta_{\text{hitting } x} = (1/2) + 1/(2^{n/2} \underline{r}(L_1)),$$
where $2^{n/2} \underline{r}(L_1)$ is exponential in $n$. It means that $1/(2^{n/2} \underline{r}(L_1))$ is a negligible function of $n$. Hence, for every (nonzero) polynomial function $poly(n)$ (notice that in the new cryptosystem, it may be a subexponential function), there exists $n_0$ such that $1/(2^{n/2} \underline{r}(L_1)) < 1/poly(n)$ for all $n > n_0$.

In summary, the new asymmetric cryptosystem is indistinguishable under chosen plaintext attack (IND-CPA). □

## 6 Conclusion

We design an asymmetric cryptosystem which can do encryption and decryption quickly by separately using a public key and a related private key.

The new asymmetric cryptosystem is analyzed to be resistant to the Shamir minima accumulation point attack and the LLL lattice basis reduction attack, and moreover proved to be semantically secure (namely IND-CPA) on the assumption that AMPP and ASSP have no subexponential time solutions.

The new cryptosystem is promised to be resistant to quantum computing attack because AMPP is equivalent to a multivariate polynomial problem, and ASSP is essentially a shortest vector problem.

Moreover, the new cryptosystem is especially suitable to the secret communication between two mobile terminals in maneuvering field operations under any weather for it has little computation and quick speed.

In future, relevant work, including analysis of the cryptosystem's non-malleability, analysis of the cryptosystem's security against adaptive chosen ciphertext attack, analysis of uniqueness of a plaintext from a ciphertext, analysis of time complexity of the algorithms, discussion of the function $\underline{r}(L_1)$, experiments on the LLL lattice basis reduction, comparison with other asymmetric cryptosystems, etc, still needs to be done.

**Acknowledgment**

The authors would like to thank the Acad. Jiren Cai, Acad. Zhongyi Zhou, Acad. Zhengyao Wei, Acad. Changxiang Shen, Acad. Yongnian Lin, Acad. Binxing Fang, Acad. Guangnan Ni, Acad. Andrew C. Yao, Acad. Jinpeng Huai, Acad. Xicheng Lu, Prof. Jie Wang, Rese. Hanliang Xu, Prof. Zhiying Wang, Prof. Ron Rivest, Prof. Moti Yung, Prof. Dingzhu Du, Prof. Ping Luo, Rese. Baodong Zhang, Rese. Shizhong Wu, Prof. Yixian Yang, Prof. Maozhi Xu, Prof. Yupu Hu, Prof. Zhiqiu Huang, Prof. Jianfeng Ma, Prof. Yongfei Han, Prof. Dingyi Pei, Prof. Mulan Liu, Prof. Lequan Min, Prof. Bogang Lin, Prof. Renji Tao, and Prof. Quanyuan Wu for their important suggestions, corrections, and helps.

## Appendix A – An Example

This example is mainly used to illustrate the correctness of the encryption and decryption of the new cryptographic scheme. The process is simplified, and a random padding is not considered.

### A.1 Key Pair

Let $n = 8$.

Generate an extra superincreasing sequence $\{A_1, \ldots, A_8\} = \{2, 4, 11, 29, 76, 199, 523, 1368\}$.

Find $M = 3581$.

Pick $\delta = 1128$ and $W = 863$.

Seek $\delta^{-1} = 1127 \% 3581$ and $(-W) = 2718 \% 3581$.

Produce $\{\ell(1), \ldots, \ell(8)\} = \{13, 2, 9, 7, 8, 3, 6, 11\}$.

Compute a sequence $\{C_1, \ldots, C_8\} =$
$\{2034, 3376, 134, 88, 2402, 746, 2833, 607\}$ by $C_i \equiv (A_i + W\ell(i))\delta \ (\% \ M)$.

Thus, a public key $(\{C_1, \ldots, C_8\}, 3581)$ and a private key $(\{A_1, \ldots, A_8\}, 1127, 2718, 3581)$ are acquired, and $\{\ell(i)\} = \{13, 2, 9, 7, 8, 3, 6, 11\}$ is discarded.

## A.2 Encryption

Assume that a public key is $(\{2034, 3376, 134, 88, 2402, 746, 2833, 607\}, 3581)$, and a plaintext block is $b_1 \ldots b_8 = 10101001$.

Yield a random noise $r_1 \ldots r_8 = 00100111$.

Compute $\{L_1, \ldots, L_8\} = \{4, 3, 3, 2, 2, 1, 1, 1\}$ by $L_i = \sum_{j=i}^{n} b_j$.

Compute a ciphertext
$$\dot{S} \equiv 1 \cdot 4 \cdot 2034 + 1 \cdot 3 \cdot 134 + 1 \cdot 2 \cdot 2402 + 1 \cdot 1 \cdot 746 + 1 \cdot 1 \cdot 2833 + 1 \cdot 1 \cdot 607 \equiv 3204 \ (\%$$
3581) by $\dot{S} \equiv \sum_{i=1}^{n} (b_i \oplus r_i) L_i C_i \ (\% \ M)$.

## A.3 Decryption

Assume that a private key is $(\{2, 4, 11, 29, 76, 199, 523, 1368\}, 1127, 2718, 3581)$, and a ciphertext is $\dot{S} = 3204$.

Compute $\dot{S} \equiv 3204 \times 1127 \equiv 1260 \ (\% \ 3581)$ by $\dot{S} \equiv \dot{S}\delta^{-1} \ (\% \ M)$.

Compute $\dot{S} \equiv 1260 + 2718 \times 115 \equiv 2283 \ (\% \ 3581)$ by $\dot{S} \equiv \dot{S} + (-W)\sum_{i=1}^{n} \ell(i)(b_i \oplus r_i) L_i \ (\% \ M)$ (note that in practice, this step is a heuristic loop due to unknown $b_i$ and $r_i$).

Set $S = \dot{S} = 2283$, and $b_1 \ldots b_8 = 0$.

Due to $2283 \geq 1368$, $S = 2283 - 1368 = 915$, $L = L + 1 = 1$, $b_8 = 1$.

Due to $523 \times L \leq 915 < 523 \times (L + 1) = 523 \times 2 = 1046$, $S = 915 - 523 = 392$, $L = 1$, $b_7 = 0$.

Due to $199 \times L \leq 392 < 199 \times (L + 1) = 199 \times 2 = 398$, $S = 392 - 199 = 193$, $L = 1$, $b_6 = 0$.

Due to $152 = 76 \times (L + 1) \leq 193$, $S = 193 - 152 = 41$, $L = L + 1 = 2$, $b_5 = 1$.

Due to $29 \times 0 < 41 < 29 \times L = 29 \times 2 = 58$, $S = 41$, $L = 2$, $b_4 = 0$.

Due to $33 = 11 \times (L + 1) \leq 41$, $S = 41 - 33 = 8$, $L = L + 1 = 3$, $b_3 = 1$.

Due to $4 \times 0 < 8 \leq 4 \times L = 4 \times 3 = 12$, $S = 8$, $L = 3$, $b_2 = 0$.

Due to $8 = 2 \times (L + 1) \leq 8$, $S = 8 - 8 = 0$, $L = L + 1 = 4$, $b_1 = 1$.

At last, a related plaintext block 10101001 is obtained.